\documentclass[journal]{IEEEtran}
\usepackage[T1]{fontenc}
\usepackage{graphicx}
\usepackage{times}
\usepackage{helvet}
\usepackage{courier}
\usepackage{amsmath}
\usepackage{algorithm}
\usepackage{algorithmic}
\usepackage{csquotes} 
\usepackage{color}
\usepackage{paralist}
\usepackage{amssymb}
\usepackage{indentfirst}
\usepackage{float}
\usepackage{multirow}
\usepackage{cite}
\usepackage{mathrsfs}

\usepackage{mathrsfs} 
\usepackage{amsfonts}
\usepackage{todonotes}
\usepackage{pgfplots} 
\usepackage{epstopdf}
\usepackage{epsfig}
\pgfplotsset{compat=newest}
\usepackage{color}
\definecolor{forestgreen}{RGB}{0,139,69}

\usepackage{xcolor}
\definecolor{citecolor}{HTML}{0071bc}
\usepackage[colorlinks, linkcolor=red,  anchorcolor=blue, citecolor=citecolor]{hyperref} 

\usepackage{xcolor}
\definecolor{SeaGreen4}{RGB}{0,205,102} 
\definecolor{SlateBlue}{RGB}{106,90,205} 
\definecolor{DarkRed}{RGB}{178,34,34} 
	
\usepackage[switch]{lineno}

\usepackage{textcomp,booktabs}
\usepackage{amssymb}
\usepackage{pifont}

\usepackage{makecell}
\usepackage{subcaption}

\usepackage{colortbl}
\definecolor{mygray}{gray}{.9}
\definecolor{mypink}{rgb}{.99,.91,.95}
\definecolor{mycyan}{cmyk}{.3,0,0,0}

\begin{document}

\title{ Fusing Global Diagnostic Signals and Spectral Data for Electron and Ion Temperature Profile Prediction on EAST via Attention-based Neural Networks } 

\author{Xiao Wang, Jiangtao Yu, Zichao Lin*, Weidong Chen, Dengdi Sun, Wanli Lyu, \\ 
    Hongming Zhang, Jin Tang*, Bo Lyu  

\thanks{$\bullet$ Xiao Wang, Jiangtao Yu, Wanli Lyu, Jin Tang are with the Anhui Provincial Key Laboratory of Multimodal Cognitive Computation, School of Computer Science and Technology, Anhui University, Hefei 230601, China (email: \{xiaowang, tangjin, lwl\}@ahu.edu.cn, yjt17879389267@foxmail.com)}  

\thanks{$\bullet$ Dengdi Sun is with School of Computer Science and Technology, Anhui University, Hefei 230601, China (email: sundengdi@ahu.edu.cn)}

\thanks{$\bullet$ Zichao Lin, Hongming Zhang, Bo Lyu are with the Institute of Plasma Physics, Hefei Institutes of Physical Science, Chinese Academy of Sciences, Hefei 230031, China (email: \{zichao.lin, hmzhang, blu\}@ipp.ac.cn)} 

\thanks{$\bullet$ Weidong Chen is with the School of Information Science and Technology, University of Science and Technology
of China, Hefei 230026, China (email: chenweidong@ustc.edu.cn)} 

\thanks{* Corresponding Author: Zichao Lin, Jin Tang} 
}

\markboth{ IEEE Transactions on ***, 2026 } 
{Shell \MakeLowercase{\textit{et al.}}: Bare Demo of IEEEtran.cls for IEEE Journals}

\maketitle

\begin{abstract}
Ion temperature $T_\mathrm{i}$ and electron temperature $T_\mathrm{e}$ are fundamental kinetic parameters for evaluating tokamak plasma confinement, turbulent transport and magnetohydrodynamic stability. Reliable prediction of their spatiotemporal profiles is essential for long-pulse steady-state operation on the Experimental Advanced Superconducting Tokamak (EAST). However, existing EAST diagnostic systems have obvious measurement limitations, including narrow applicable conditions of charge exchange recombination spectroscopy and severe signal degradation of edge temperature diagnostics during long discharges. Classical physics-based transport codes suffer from high computational cost and error accumulation, while existing data-driven models only separately predict $T_\mathrm{i}$ or $T_\mathrm{e}$ relying on spectral data, lacking joint $T_\mathrm{i}/T_\mathrm{e}$ modeling and failing under missing spectral measurements. 
In this work, we propose a multi-modal (i.e., global macroscopic diagnostic signals and spectral data) fusion attention neural network for simultaneous reconstruction of $T_\mathrm{i}/T_\mathrm{e}$ profiles on EAST. It leverages an adaptive attention module to weight global diagnostic features and suppress noise interference, employs convolutional layers to extract deep spectral representations, and utilizes multi-head self-attention to facilitate cross-modal feature interaction. Two separate prediction heads are constructed to output full radial distributions of \(T_\mathrm{i}\) and \(T_\mathrm{e}\). 
This work builds the first EAST-oriented joint $T_\mathrm{i}/T_\mathrm{e}$ prediction model compatible with operating scenarios without spectral diagnostics. Hierarchical attention mechanisms excavate nonlinear multi-modal correlations to realize high-performance dual-temperature profile prediction.
The source code of this paper is available on \url{https://github.com/Event-AHU/OpenFusion/tree/main/TiTe_Prediction} 
\end{abstract}


\IEEEpeerreviewmaketitle

\section{Introduction}

Ion temperature ($T_\mathrm{i}$), electron temperature ($T_\mathrm{e}$), and their ratio $T_\mathrm{i}/T_\mathrm{e}$ are essential kinetic parameters that characterize the confinement performance, microscopic turbulent transport, and macroscopic magnetohydrodynamic (MHD) stability of tokamak plasmas. Accurate prediction of the temporal and spatial evolution of $T_\mathrm{i}/T_\mathrm{e}$ serves as a fundamental prerequisite for plasma diagnosis, physical analysis, and real-time control of long-pulse steady-state discharges on the Experimental Advanced Superconducting Tokamak (EAST)~\footnote{\url{http://east.ipp.ac.cn/}}. Physically, turbulent transport dominated by ion temperature gradient (ITG) and electron temperature gradient (ETG) instabilities, as well as neoclassical transport behaviors under diverse EAST operational regimes, strongly depend on the radial gradients of $T_\mathrm{i}$, $T_\mathrm{e}$, and the $T_\mathrm{i}/T_\mathrm{e}$ ratio. 
Furthermore, the dynamic evolution of the $T_\mathrm{i}/T_\mathrm{e}$ profile acts as a critical precursor for the excitation and development of MHD instabilities, including sawtooth oscillations, neoclassical tearing modes (NTMs), and edge-localized modes (ELMs), enabling advanced disruption prediction and active stability regulation. 
Several diagnostic systems have been developed for temperature measurements on EAST. Charge exchange recombination spectroscopy (CXRS)~\cite{Li2014EASTCXRS} and X-ray crystal spectroscopy~\cite{Lyu2014EASTXICS} are mainly used for $T_\mathrm{i}$ measurements, whereas Thomson scattering (TS)~\cite{Zang2011EASTTS} and electron cyclotron emission (ECE)~\cite{Zhao2018EASTECE} diagnostics are primarily used for $T_\mathrm{e}$ measurements. However, these dedicated temperature diagnostics are not available for every discharge or throughout all phases of a discharge because of their specific operational requirements, limited measurement coverage, diagnostic faults, or insufficient data quality. As a result, complete $T_\mathrm{i}$ and $T_\mathrm{e}$ profiles are unavailable for a considerable number of EAST discharges, which limits temperature-related statistical analyses and comparisons among different plasma scenarios. 

\begin{figure*}
\centering
\includegraphics[width=1\textwidth]{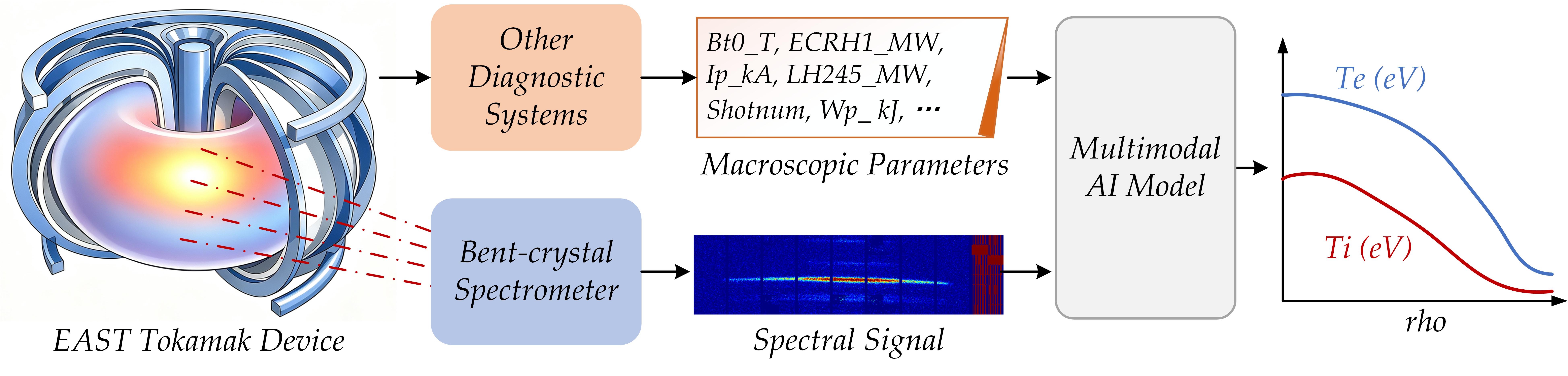}
\caption{Schematic overview of the multimodal AI‑driven plasma temperature prediction framework on the EAST tokamak.} 
\label{fig:firstIMG}
\end{figure*}

In contrast, global plasma parameters, including the plasma current, stored energy, and auxiliary heating powers, are routinely measured and are generally available for a much larger number of discharges. These macroscopic parameters contain information associated with the plasma heating conditions, energy confinement, and overall discharge state. Establishing a data-driven relationship between these routinely available parameters and the temperature profiles therefore provides a practical approach for estimating $T_\mathrm{i}$, $T_\mathrm{e}$, and $T_\mathrm{i}/T_\mathrm{e}$ when direct temperature measurements are missing or incomplete. Conventional fitting and inversion methods~\cite{cenacchi1988jetto,hawryluk1979empirical,felici2018real} for estimation of $T_e$ and $T_i$ are constrained by limited diagnostic accuracy and high computational cost.However, model-based $T_\mathrm{i}/T_\mathrm{e}$ prediction can effectively reconstruct complete temperature profiles, compensate for the deficiencies of hardware-based measurements, and reduce the computational cost of steady-state plasma diagnostics.
Consequently, research on accurate $T_\mathrm{i}/T_\mathrm{e}$ prediction is of significant theoretical and engineering importance for improving the diagnostic framework of EAST, revealing the underlying physics of steady-state plasma transport and stability, and advancing long-pulse operation technologies for future fusion reactors.


Data-driven approaches have become an important trend in the inference of plasma kinetic profiles, including $T_e$ and $T_i$~\cite{morosohk2025estimation}. Previous studies mainly focus on predicting the temporal evolution of plasma kinetic profiles, thereby providing a methodological foundation and experimental support for the joint prediction of $T_e$ and $T_i$. More in detail, previous studies~\cite{lu2023observations, lyu2014upgrades, wang2016upgrades} on EAST have demonstrated that XCS-based spectral data can be used to infer both ion and electron temperature profiles on EAST. In particular, data-driven models have been developed for $T_i$ prediction directly from raw XCS spectra, while a convolutional neural network (CNN) model~\cite{lin2025development, lin2024prediction} has also been proposed for $T_e$ reconstruction from XCS-related impurity emission profiles. These results confirm the feasibility of spectrum-based temperature profile inference. However, they still rely strongly on spectral measurements and do not provide a joint prediction of $T_e$ and $T_i$. Besides, they are not applicable when spectral data are unavailable; in such cases, conventional studies often rely on simulation tools to provide self-consistent profile estimates. In conventional studies, temperature prediction typically relies on physics-based integrated modeling frameworks, such as JETTO~\cite{cenacchi1988jetto}, TRANSP~\cite{hawryluk1979empirical}, and RAPTOR~\cite{felici2018real}, especially when direct profile measurements are unavailable~\cite{waltz1997gyro, waltz2013tglf, meneghini2015omfit}. These codes simulate profile evolution by coupling multiple physical modules. 
Although they provide physically self-consistent results, they depend on strong assumptions regarding turbulence, transport, and equilibrium, which are prone to error accumulation and incur high computational costs. As a result, they are not well-suited to high-frequency real-time applications. Data-driven methods have provided an effective way to overcome these constraints~\cite{eldon2025latent}. Abbate et al. proposed a fully data-driven CNN–LSTM hybrid model for plasma profile evolution prediction in DIII-D~\cite{abbate2021data}. The model captured spatial features via CNN and temporal dependencies via LSTM, realizing fast forward prediction of $T_e$ profiles. However, this work aimed at time-series prediction and did not include $T_i$ nor involve simultaneous $T_e$ and $T_i$ estimation. 
In summary, data-driven methods have shown distinct advantages in plasma temperature prediction. However, simultaneous and robust prediction of $T_e$ and $T_i$ remains insufficiently investigated. A dedicated model that jointly learns the nonlinear correlation between $T_e$ and $T_i$ and adapts to full discharge scenarios is still highly desirable.

To address the aforementioned challenges, this paper proposes a novel multi-modal fusion-driven framework for Ti/Te profile evolution, as illustrated in Figure~\ref{fig:firstIMG}. Specifically, given the global diagnostic signals and XCS spectral data, the proposed framework first preprocesses the macroscopic parameters. A dimension expansion operation is adopted to enhance the feature representation capability of input signals, and multiple linear layers are then utilized to extract effective features from the preprocessed data. Considering that different diagnostic signals carry distinct information quantities, an attention mechanism is designed to adaptively weight various diagnostic signals. This strategy enables the model to assign higher attention weights to critical diagnostic signals and reduce the adverse impact of interfering inputs. For the two-dimensional spectral data, a convolutional neural network is introduced to extract deep spectral features, and linear layers are employed for feature mapping. The mapped spectral features are further concatenated with the weighted diagnostic signal features and are fed into a multi-head self-attention network to achieve deep interaction and fusion of cross-modal features. Finally, two independent Ti/Te prediction heads are established to realize high-precision estimation of profile evolution.

To sum up, the main contributions of this paper can be summarized as the following three aspects: 
\textbf{(1).} We propose the first electron and ion temperature prediction framework for the EAST tokamak, which enables the reconstruction of electron and ion temperature profiles using global plasma parameters, even in the absence of spectral diagnostic signals. 
\textbf{(2).} It leverages effective attention weighting and self-attention mechanisms to learn discriminative multi-modal features from global diagnostic signals and spectral data, thereby achieving high-precision prediction of $T_i/T_e$.  
\textbf{(3).} Extensive experiments fully validated the effectiveness of our proposed attention-based neural networks for the electron and ion temperature prediction.

\textit{The rest of this paper is organized as follows:}  
In Section~\ref{sec::relatedWorks}, we give an introduction to Electron and Ion Temperature Prediction, Multi-modal Fusion, and Attention Mechanism. Then, we describe the key procedures of our proposed framework in Section~\ref{sec::method}. After that, we conduct the experiments to validate the effectiveness in Section~\ref{sec::Experiments} and conclude this paper in Section~\ref{sec::Conclusion}, respectively.

\section{Related Works} \label{sec::relatedWorks}

In this section, we will introduce the related works on artificial intelligence techniques, i.e., multi-modal fusion and attention mechanism.

\subsection{Multi-modal Fusion} 
Multi-modal fusion aims to integrate complementary information from heterogeneous data sources to improve the representation ability, robustness, and generalization performance of data-driven models~\cite{baltruvsaitis2018multimodal,jiao2024comprehensive, stahlschmidt2022multimodal,guo2019deep,liang2024foundations,zhang2024multimodal}. In tokamak plasma diagnostics, input signals show obvious heterogeneous structural characteristics, mainly including zero-dimensional macroscopic scalar parameters and high-dimensional two-dimensional spectral data~\cite{lin2024prediction, ling2025pamma}. These two types of signals describe the plasma state from global equilibrium and local diagnostic details respectively, and are highly complementary in information coverage~\cite{wang2025physics}.

With the rapid development of deep learning, feature-level fusion frameworks based on neural networks have become the mainstream of multi-modal learning~\cite{jalalvand2025multimodal,poels2025plasma,seo2023multimodal}. Such frameworks encode different modalities through independent branches, project heterogeneous features into a unified latent space, and then realize adaptive information fusion via attention mechanisms. It can not only retain the unique representation of each modality, but also deeply mine cross-modal correlations and global joint information.
In the joint prediction of electron and ion temperature profiles, most existing models rely on single-modal input~\cite{abbate2021data} and lack an effective fusion scheme for global diagnostic signals and local spectral features. Although multi-modal fusion technology has made progress in heterogeneous information integration, a dedicated fusion architecture for EAST electron and ion temperature joint prediction, which is lightweight and physically constrained, is still lacking. Such an architecture should effectively couple low-dimensional global features and high-dimensional spectral structural features, and provide high-quality fused feature representations for the subsequent spatial modeling of temperature profiles.

\subsection{Attention Mechanism}
Attention mechanisms originate from the principle of human visual perception~\cite{itti1998model}. Their core objective is to enable models to assign weights to input information~\cite {bahdanau2014neural,luong2015effective,hu2018squeeze,li2019selective,woo2018cbam,dao2022flashattention}, focus on key data, suppress redundancy, and enhance feature mining and information screening capabilities.
Attention networks can adaptively generate weight coefficients based on input data characteristics. By performing a weighted fusion with original features, they optimize feature representation and address the limitations of traditional networks in extracting global features uniformly.
The Transformer model~\cite {vaswani2017attention}, based on the self-attention mechanism, has become the mainstream framework in current time-series modeling~\cite {wu2021autoformer,zhou2022fedformer,nie2022time,zhang2023crossformer}. It abandons the sequential processing of traditional recurrent networks~\cite {rumelhart1986learning,graves2012long,cho2014learning}, efficiently captures long-range dependencies through multi-head self-attention, supports parallel computing, and significantly improves both training efficiency and global feature modeling.
Transformers and their derived attention networks~\cite {zhou2021informer,beltagy2020longformer,liu2021swin} have been widely applied to tasks such as time-series analysis and multi-source information fusion. While existing studies predominantly utilize attention structures to optimize feature extraction, lightweight attention architectures tailored for the collaborative modeling of multi-source heterogeneous data in tokamak plasmas require further refinement.

\section{Our Proposed Method} \label{sec::method}

\begin{figure*}[htbp] 
\centering     
\includegraphics[width=\textwidth]{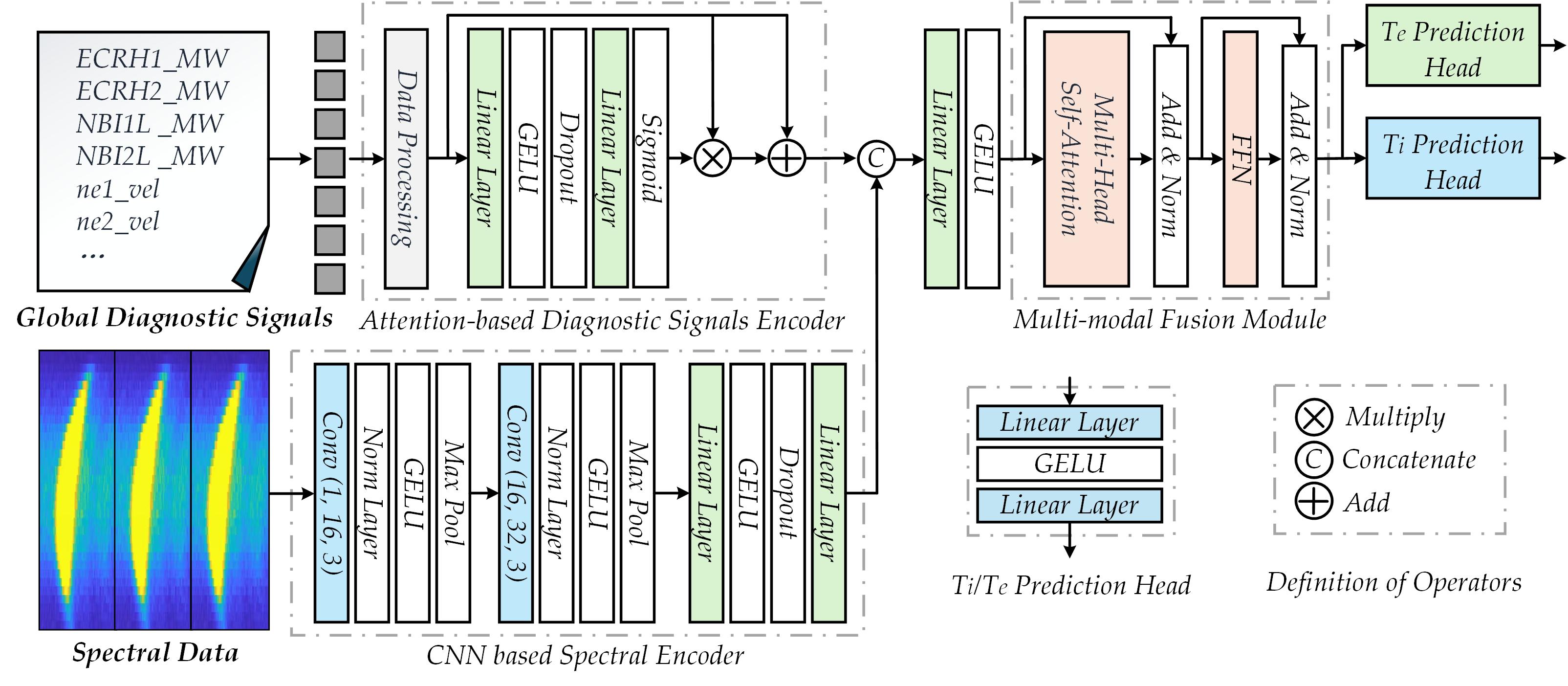}
\caption{An overview of our proposed attention-based neural networks for $T_i/T_e$ profile prediction using global diagnostic signals and spectral data. It adopts an attention-based diagnostic signal encoder and a CNN-based spectral encoder to perceive the given multi-modal information and fuse them using the multi-head self-attention layers. Two independent task heads are utilized to predict the evolution profiles of $T_i$ and $T_e$, respectively. The proposed method can effectively encode multimodal information for discriminative feature learning and data fusion. Under missing spectral conditions, accurate temperature profile estimation can be achieved relying merely on global diagnostic signals. 
}  
\label{fig:framework} 
\end{figure*}

\subsection{Problem Formulation} 
Traditional physics-based models rely on simplified transport assumptions and iterative numerical calculations, which introduce error accumulation and high computational latency. Existing data-driven methods only utilize spectral information for separate \(T_i\) or \(T_e\) estimation and fail to model the inherent coupling relationship between ion and electron temperature evolution. Such methods cannot maintain stable prediction performance when spectral data are missing. In this section, we formally formulate a multi-modal fusion-based joint profile prediction task, which takes global diagnostic signals $X_g$ and spectral measurements $X_s$ as heterogeneous input modalities. 
The core task is to establish a nonlinear spatiotemporal mapping function $F(\cdot)$ that projects multi-modal input features to complete $T_i$ and $T_e$ radial temperature distributions, which can be mathematically expressed as: 
\begin{equation}
    \label{} 
    [T_i(\rho,t),~T_e(\rho,t)] = F(X_g,~X_s) 
\end{equation} 
where $\rho \in [0,1]$ represents the normalized radial coordinate of the tokamak plasma cross section, and $t$ denotes the discharge time step. The model outputs discretized radial profiles of both $T_i$ and $T_e$ on a fixed grid of 32 points over $\rho \in [0,1]$, yielding a total output dimensionality of 64. 
The trained model is expected to produce high-fidelity dual-temperature profile reconstruction under normal conditions with complete spectral data and maintain stable generalization performance in scenarios where spectral measurements are unavailable. It provides reliable profile estimation to support steady-state plasma physical analysis and real-time discharge regulation for EAST long-pulse operations.

\subsection{Overview} 
As shown in Figure~\ref{fig:framework}, this paper proposes an attention-based multi-modal Transformer learning framework that fuses global diagnostic signals and spectral data for the joint prediction of electron temperature $T_e$ and ion temperature $T_i$ profiles in EAST tokamak plasmas. Two types of complementary heterogeneous modal data are taken as the input, including the global diagnostic signals $X_g$ and the spectral diagnostic data $X_s$. Traditional CNNs mainly focus on local receptive fields and are insufficient to capture the globally correlated structural information in spectral data, making it difficult to fully characterize the global morphology and coupling laws of temperature profiles along the radial direction. 
In contrast, Transformers have been widely proven to have significant advantages in modeling long-range dependencies and global structures. 
Therefore, this paper adopts the Transformer architecture in the multi-modal feature fusion stage to perform global modeling and interactive coding on the features of spectral modality and macroscopic temporal diagnostic modality, so as to obtain a unified feature representation with global context information. 

On this basis, the model constructs radial sequence representations through spatial tokenization and positional embedding, captures the long-range spatial dependencies and core-edge coupling relationships of temperature profiles along the radial direction using the Transformer fusion network, and adopts a dual-prediction-head structure to achieve the simultaneous output of $T_e$ and $T_i$ profiles. This design forms a unified learning architecture with multimodal collaboration, global perception, and spatial structuring, which can well adapt to plasma temperature profile reconstruction in real tokamak experimental environments, providing a feasible solution for efficient and accurate joint prediction of electron and ion temperatures.

\begin{table}[htbp]
\centering
\caption{Tokamak Plasma Diagnostic Parameters Description}
\label{tab:plasma_parameters}
\resizebox{\linewidth}{!}{
\begin{tabular}{ll} 
    \toprule
    \textbf{Name} & \textbf{Physical / Practical meaning} \\
    \midrule
    Bt0\_T & Toroidal magnetic field strength at the magnetic axis, in Tesla \\
    ECRH1\_MW & Injected power of ECRH system/channel 1, in MW \\
    ECRH2\_MW & Injected power of ECRH system/channel 2, in MW \\
    ECRH3\_MW & Injected power of ECRH system/channel 3, in MW \\
    ECRH4\_MW & Injected power of ECRH system/channel 4, in MW \\
    ICRH\_MW & Injected power of ion cyclotron resonance heating, in MW \\
    Ip\_kA & Plasma current, in kA \\
    LH245\_MW & Injected power of 2.45 GHz lower hybrid wave heating, in MW \\
    LH460\_MW & Injected power of 4.60 GHz lower hybrid wave heating, in MW \\
    NBI1L\_MW & Injected power of neutral beam injection source 1-left, in MW \\
    NBI1R\_MW & Injected power of neutral beam injection source 1-right, in MW \\
    NBI2L\_MW & Injected power of neutral beam injection source 2-left, in MW \\
    NBI2R\_MW & Injected power of neutral beam injection source 2-right, in MW \\
    Shotnum & Discharge shot number \\
    Spec & Measured spectral data \\
    Te\_out & Electron temperature profile \\
    Ti\_out & Ion temperature profile \\
    Wp\_kJ & Plasma stored energy, in kJ \\
    drsep & Distance between the primary and secondary separatrices at the outer midplane \\
    li\_val & Plasma internal inductance \\
    ne1\_val & Electron density from the solid-source diagnostic \\
    ne2\_val & Electron density from the POINT diagnostic \\
    q0\_val & Safety factor at the magnetic axis \\
    q95\_val & Safety factor at 95\% normalized poloidal flux surface \\
    t & Time point or time array \\
    \bottomrule
  \end{tabular}} 
\end{table}

\subsection{Input Representation} 


In our framework, the macro branch $X_g$ is a 20-dimensional vector composed of scalar discharge descriptors excluding metadata and target fields, as fully listed in Table~\ref{tab:plasma_parameters}. Before standardization, a noise filtering procedure is implemented to eliminate outliers and random measurement artifacts from raw macroscopic diagnostic signals, which alleviates the adverse impact of abnormal noisy samples on subsequent model training. These cleaned features are then standardized using the mean and standard deviation computed on the training set. After being processed by the attention-based diagnostic signals encoder, the macro branch finally outputs the feature vector $X_m$, which retains the original dimension of 20.

In parallel, the spectral branch is constructed from the raw diagnostic matrix. The original spectral tensor is first reshaped, then cropped according to the discharge-dependent spectral window, and finally aggregated over 35 predefined intervals along one dimension. After the above cropping and aggregation operations, the processed spectral tensor yields a fixed shape of $35 \times 66$ and is denoted as $X_{spec}$. This representation preserves structured spectral information while substantially reducing dimensionality. $X_{spec}$ is also normalized using training-set statistics.

This dual-stream design is motivated by the complementary nature of the inputs. The macro variables provide global operating-state information, such as plasma equilibrium and actuator conditions, whereas the spectral signal carries richer localized diagnostic structure. A successful predictor can fuse low-dimensional physics descriptors with high-dimensional measurement patterns in a unified latent space. After concatenating the weighted macroscopic features $X_m$ and compact spectral representation $X_{spec}$, a dimension expansion operation is further applied to the fused latent vector, so as to match the feature dimension required by the follow-up multi-modal fusion module.

\subsection{Network Architecture} 

As shown in Figure~\ref{fig:framework}, our model consists of three main components, i.e., an Attention-based Diagnostic Signals Encoder, a CNN-based Spectral Encoder, and a Multi-modal Fusion Module. It achieves deep fusion of heterogeneous diagnostic information and high-precision modeling of temperature profiles.

\noindent $\bullet$ \textbf{Attention-based Diagnostic Signals Encoder.} 
The macroscopic diagnostic signals consist of a set of low-dimensional scalar parameters describing the plasma operating state, including key physical quantities, such as toroidal magnetic field, heating power, plasma current, safety factor, and confinement parameters. These parameters contribute significantly differently to the temperature profiles, and direct simple concatenation will introduce redundant information and disturb model convergence.

To handle these diagnostic signals, a lightweight attention-based network is designed to achieve adaptive weighting and enhancement of diagnostic features in a data driven manner. The module takes the standardized macroscopic feature vector as input, and consists of two fully connected layers (${FC}_1, {FC}_2$), nonlinear activation {GELU}, dropout regularization, and a sigmoid gating function $\sigma(*)$. The module first performs a nonlinear transformation on the input macroscopic features to generate an attention weight vector $w$ with the same dimension as the input, and then normalizes the weights to the $[0,1]$ interval via the sigmoid function to characterize the relative importance of each channel feature. This process can be briefly formulated as: 
\begin{equation}
w = \sigma({GELU} ({FC}_2({GELU}({FC}_1(x_m)))))
\end{equation}

The final output adopts a residual fusion structure:
\begin{equation}
\tilde{x}_m = x_m + x_m \odot w
\end{equation}
where $x_m$ is the original macroscopic feature, and $\odot$ denotes element wise multiplication. The residual structure can enhance critical diagnostic information while retaining the complete information of the original physical features, avoiding information loss caused by hard selection, and significantly improving the robustness and prediction stability of the model under complex discharge conditions.

\noindent $\bullet$ \textbf{CNN based Spectral Encoder.} 
This paper constructs a dedicated two-dimensional spectral convolutional backbone network to map high-dimensional spectral data into compact and discriminative embedding vectors. The backbone is composed of two consecutive convolutional blocks. Each block sequentially includes a two-dimensional convolutional layer, a batch normalization layer, a GELU activation function, and an average pooling layer. The convolutional layer is used to extract local textures and spatial correlations of spectral data; batch normalization accelerates network convergence and improves training stability; GELU activation provides stronger nonlinear representation capability; and the pooling layer gradually reduces the feature scale, retains key information and suppresses noise. This process can be written as: 
\begin{align}
\boldsymbol{F}_{l} &= \text{AvgPool}\big(\text{GELU}\big(\text{BN}\big(\text{Conv2d}\big(\boldsymbol{F}_{l-1}\big)\big)\big)\big), l=\{1,2\} \\
\boldsymbol{F}_{0} &= \boldsymbol{X}_{\text{spec}}
\end{align}

After being extracted by the convolutional backbone, the output three-dimensional feature map is flattened into a one-dimensional vector $\boldsymbol{f}_{{flat}} = {Flatten}(\boldsymbol{F}_2)$, and then mapped into a fixed-dimensional spectral embedding via a multilayer perceptron with batch normalization and dropout, i.e., $\boldsymbol{e}_s = {MLP}(\boldsymbol{f}_{{flat}})$. This process preserves the global structural information of the spectrum while achieving efficient compression from high-dimensional data to low-dimensional embedding, providing a unified and compact representation basis for the subsequent cross-modal fusion with macroscopic features.

\noindent $\bullet$ \textbf{Multi-modal Fusion Module.~} 
The attended macro representation {$\tilde{\boldsymbol{x}}_m$} and the spectral embedding {$\boldsymbol{e}_s$} are concatenated into a fused latent vector {$\tilde{\boldsymbol{X}}$}. Instead of directly regressing the full temperature profile from this vector, the fused representation is projected into a sequence of spatial tokens with shape $N_{\rho} \times d$, where $N_{\rho}=32$ is the number of radial grid points and $d$ is the embedding dimension. In this way, each radial location is associated with a learnable token representation before contextual refinement. The resulting sequence, denoted as \(\mathbf{X}\), serves as the input to the subsequent Transformer encoder after adding the positional embeddings. Specifically, let \(\mathbf{E}_{{pos}} \in \mathbb{R}^{N_\rho \times d}\) denote the learnable positional embedding, and we have \(\mathbf{Z}_0 = \mathbf{X} + \mathbf{E}_{{pos}}\).
For simplicity, we omit the symbol \(\mathbf{Z}_0\) and use \(\mathbf{X}\) to represent the position-augmented sequence in the following attention formulas.

The spatial token sequence is subsequently passed through a multi-layer transformer encoder. For each attention head, the input token sequence is linearly projected into query, key, and value matrices:
\begin{equation}
\mathbf{Q}_h = \mathbf{X}\mathbf{W}_h^Q,\qquad
\mathbf{K}_h = \mathbf{X}\mathbf{W}_h^K,\qquad
\mathbf{V}_h = \mathbf{X}\mathbf{W}_h^V,
\end{equation}
where $\mathbf{W}_h^Q,\mathbf{W}_h^K,\mathbf{W}_h^V$ are learnable projection matrices and $h$ denotes the attention head. 
The scaled dot-product self-attention for the $h$-th head is defined as:
\begin{equation}
\mathrm{SelfAtten}_h(\mathbf{X})=\mathrm{Softmax}\!\left(\frac{\mathbf{Q}_h \mathbf{K}_h^{\top}}{\sqrt{d_h}}\right)\mathbf{V}_h,
\end{equation}
where $d_h$ is the feature dimension of each attention head. The outputs of all heads are concatenated and projected to produce the multi-head self-attention output:
\begin{equation}
\small 
\mathrm{MHSA}(\mathbf{X})=\mathrm{Concat}\!\left(\mathrm{SelfAtten}_1(\mathbf{X}), \dots, \mathrm{SelfAtten}_H(\mathbf{X})\right)\mathbf{W}^O.
\end{equation}

Through this mechanism, each radial token interacts with all other radial positions, thereby capturing profile-wide dependencies and global shape constraints. This design is more suitable than purely local operators when the profile morphology depends on nonlocal transport behavior or coordinated core-edge changes.

After transformer encoding, the refined spatial tokens are fed into two independent prediction heads for $T_i$ and $T_e$, respectively. Each head is instantiated as a lightweight multilayer perceptron that operates point‑wise on the transformer output tokens: 
\begin{equation}
\hat{\mathbf{T}}_i = MLP_{ti}(\mathbf{X}^{(L)}),\qquad
\hat{\mathbf{T}}_e = MLP_{te}(\mathbf{X}^{(L)}),
\end{equation}
where $MLP_{ti}(\cdot)$ and $MLP_{te}(\cdot)$ denote two independent MLP heads. This dual-head design is particularly important because ion and electron temperatures are correlated but not identical physical quantities. By decoupling the final regression stage, the model reduces gradient interference between the two tasks while still allowing them to share a common multimodal and spatially aware backbone.

The proposed architecture offers several advantages for tokamak temperature-profile prediction. 
\textit{First}, it explicitly fuses heterogeneous diagnostics rather than forcing all inputs into a single undifferentiated branch. 
\textit{Second}, the macro-feature attention module provides adaptive reweighting of scalar diagnostics, which is both performance-enhancing and interpretable. 
\textit{Third}, the transformer backbone models long-range radial dependencies, making it more appropriate for structured profile prediction than purely pointwise regressors. 
\textit{Fourth}, the dual-head output design acknowledges the partially shared yet distinct physics of $T_i$ and $T_e$. 
\textit{Finally}, the physics-guided loss embeds domain knowledge directly into the optimization objective, improving both numerical accuracy and physical plausibility.
Taken together, these design choices define a multimodal, physics-aware, and spatially structured learning framework that is well matched to the demands of plasma profile reconstruction in realistic tokamak scenarios.

\subsection{Loss Function} 

For the optimization of our framework, this study designs a physics-inspired joint loss function that incorporates the transport characteristics of tokamak plasmas, instead of using a simple mean squared error (MSE) loss. For each task $T_i/T_e$, the overall loss function consists of three components: a weighted MSE term, a first-order spatial gradient smoothing penalty, and a physical monotonicity constraint. In addition, considering the differences in magnitude and prediction difficulty between the electron temperature ($T_e$) and ion temperature ($T_i$) profiles, an additional amplification weight is assigned to the $T_e$ branch in the final loss aggregation.

Let $s \in \{T_i, T_e\}$ denote the ion-temperature and electron-temperature branches, $T^{(s)}_{n,j}$ and $\hat{T}^{(s)}_{n,j}$ denote the ground-truth and predicted temperature values of the $n$-th sample at the $j$-th radial position, respectively. To emphasize the high-temperature core region, a radial weight is defined as: 
\begin{equation}
w_j = 
\begin{cases}
3.0, & \rho_j < 0.4,\\
1.0, & \rho_j \geq 0.4.
\end{cases}
\end{equation}
Accordingly, the Weighted MSE (WMSE) term is formulated as: 
\begin{equation}
\mathcal{L}^{(s)}_{WMSE} = \frac{1}{N} \sum_{n=1}^{N}\sum_{j=1}^{J} w_j \left(\hat{T}^{(s)}_{n,j} - T^{(s)}_{n,j}\right)^2.
\end{equation}

To enforce spatial smoothness and preserve the radial profile morphology, a first-order gradient consistency penalty is introduced. 
We define the radial finite differences as: 
\begin{equation}
\Delta \hat{T}^{(s)}_{n,j} = \hat{T}^{(s)}_{n,j+1} - \hat{T}^{(s)}_{n,j},
\qquad
\Delta T^{(s)}_{n,j} = T^{(s)}_{n,j+1} - T^{(s)}_{n,j}.
\end{equation}
Thus, the gradient smoothing loss is written as: 
\begin{equation}
\mathcal{L}^{(s)}_{\mathrm{Gradient}}=\frac{1}{N}\sum_{n=1}^{N}\sum_{j=1}^{J-1}\left(\Delta \hat{T}^{(s)}_{n,j} - \Delta T^{(s)}_{n,j}\right)^2.
\end{equation}

Furthermore, based on the physical prior that tokamak temperature profiles generally decrease monotonically from the plasma core to the edge, a monotonicity penalty is imposed on locally increasing predictions. This term is defined as: 
\begin{equation}
\mathcal{L}^{(s)}_{\mathrm{Monotonicity}}=\frac{1}{N}\sum_{n=1}^{N}\sum_{j=1}^{J-1}\mathrm{ReLU}\!\left(\Delta \hat{T}^{(s)}_{n,j}\right).
\end{equation}
Thus, the branch-wise loss for each temperature profile is expressed as: 
\begin{equation}
\mathcal{L}_{s}=\mathcal{L}^{(s)}_{WMSE}+\lambda_{1}\mathcal{L}^{(s)}_{\mathrm{Gradient}}+\lambda_{2}\mathcal{L}^{(s)}_{\mathrm{Monotonicity}},~s \in \{T_i, T_e\},
\end{equation}
where $\lambda_1 = 0.5$ and $\lambda_2 = 0.1$ are the weighting coefficients for the gradient smoothing term and the monotonicity constraint, respectively. 
As our framework handles the prediction of both $T_i$ and $T_e$, and the electron temperature profile prediction suffers from larger magnitude and higher prediction difficulty, the total loss is formulated as: 
\begin{equation}
\mathcal{L}_{\mathrm{total}}=\mathcal{L}_{Ti}+\omega \mathcal{L}_{Te},
\end{equation}
where $\omega = 2.0$ is the relative amplification factor for the $T_e$ branch.



\begin{figure}[t]
    \centering
    \includegraphics[width=\linewidth]{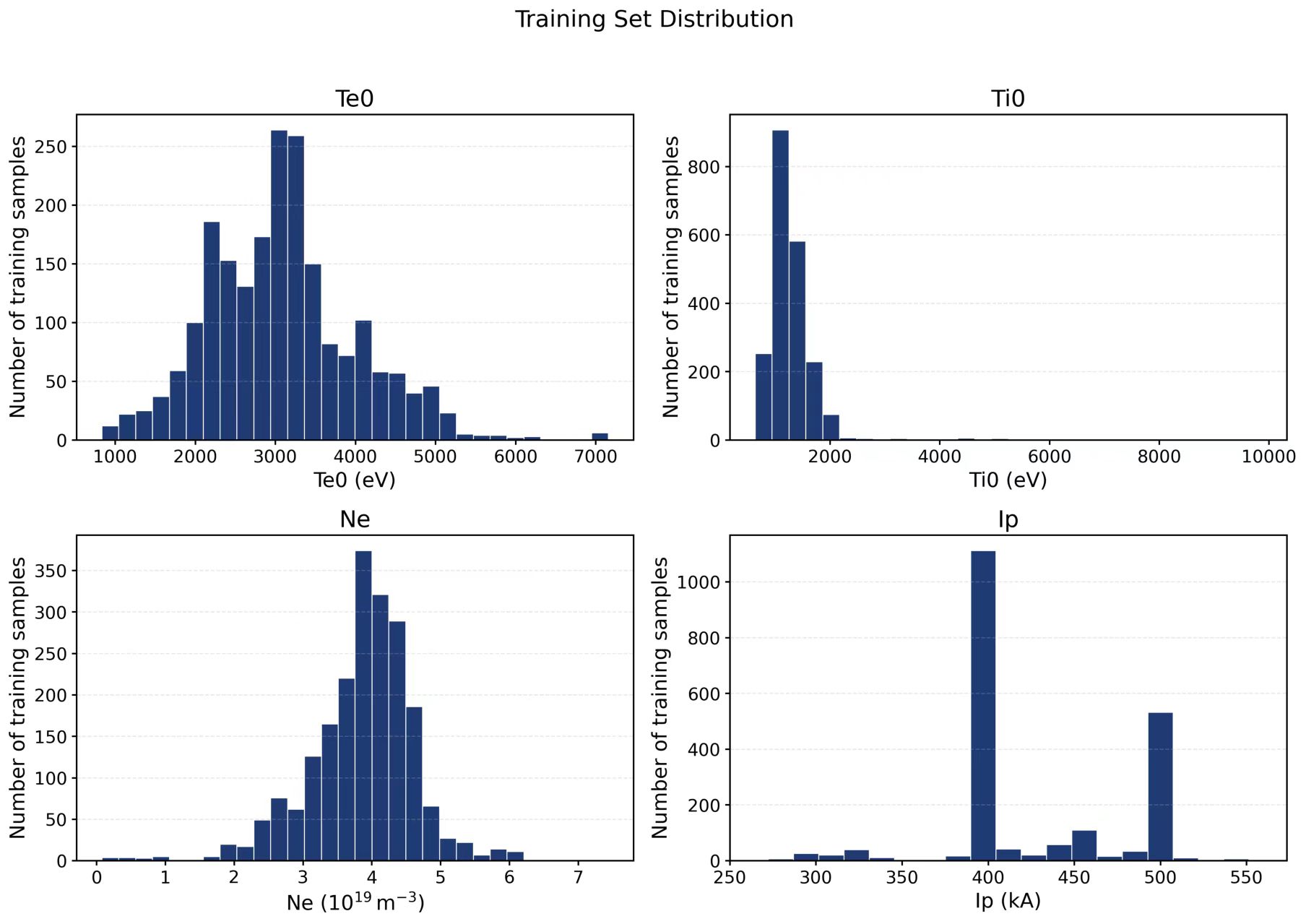}
    \caption{
    Distribution of representative physical quantities in the training set.
    The four panels show the central electron temperature $T_{e0}$,
    central ion temperature $T_{i0}$, electron density $n_e$, and plasma
    current $I_p$, respectively.
    }
    \label{fig:training_distribution}
\end{figure}

\section{Experiments} \label{sec::Experiments}

\subsection{Dataset and Evaluation Metric} 

\noindent \textbf{Dataset Information:} 
EAST shots from the \#102169(2021) to \#141698(2024) campaigns are collected from the MDS + database. Shots with a pulse length less than 2 s, RMP applied, no $T_\mathrm{i}$ or $T_\mathrm{e}$ measurement are excluded. 
To further characterize the statistical properties of the dataset, 
Figure~\ref{fig:training_distribution} shows the distributions of several 
representative physical quantities in the training set, including the central 
electron temperature $T_{e0}$, central ion temperature $T_{i0}$, electron 
density $n_e$, and plasma current $I_p$. The samples cover a relatively broad 
range of plasma conditions. In particular, $T_{e0}$ exhibits a broad 
distribution over different temperature regimes, whereas $T_{i0}$ is more 
concentrated in the lower-temperature region with a relatively sparse 
high-temperature tail. The electron density is mainly distributed around the 
typical operating range of EAST discharges, while the plasma current exhibits 
several distinct operating regimes, with pronounced concentrations around 
400~kA and 500~kA. These distributions demonstrate that the training set 
contains diverse plasma operating conditions while also exhibiting the 
non-uniform sampling characteristics inherent in experimental discharge data.

Besides, we only include data during plasma current flattop. All signals are then put on the same 200 ms time base by averaging all signal values available between the present time and 200 ms prior, based on the time resolution of XCS. 
The dataset employed in this study is derived from tokamak plasma discharge experiments. To achieve precise prediction of plasma temperature profiles, the model inputs integrate zero-dimensional (0-D) macroscopic control parameters (such as auxiliary heating power and plasma density) with two-dimensional (2D) spectral diagnostic feature data. Regarding the construction of target variables, the raw ion temperature ($T_i$) and electron temperature ($T_e$) diagnostic data appear as irregular discrete spatial samples, accompanied by varying degrees of measurement noise and missing values. Therefore, rigorous physical preprocessing was first applied to the raw discrete data. To address data sparsity in the core and edge regions, interpolation and local smoothing spline algorithms were adopted, and the latter half of the $T_i$ profile was stitched using the $T_e$ profile~\cite{meyer2011h,berkery2021kinetic}. Combined with physical monotonicity constraints, the $T_i$ and $T_e$ profiles were uniformly mapped to 32 uniform spatial grid points over the normalized poloidal flux coordinate (or normalized minor radius) $\rho \in [0, 1]$. Furthermore, to eliminate discrepancies in units and magnitudes among multi-source heterogeneous diagnostic data and to improve the convergence speed and stability of the neural network, all input features and output targets were standardized before training. The dataset contains a total of 2104 plasma discharge shots. The entire dataset was randomly partitioned into training, validation, and test sets according to the shot numbers to fully verify the model’s generalization ability. The detailed description of all variables used in this study is presented in Table~\ref{tab:plasma_parameters}. 

\noindent \textbf{Evaluation Metrics:} 
To objectively and quantitatively assess the prediction performance of deep learning models for plasma temperature profiles, this study adopts three commonly used statistical error metrics: Mean Absolute Error (MAE), Root Mean Squared Error (RMSE), and the Coefficient of Determination ($R^2$). Mean Absolute Error (MAE) measures the average magnitude of absolute errors between the model’s predicted values and the ground-truth values, and intuitively reflects the overall average deviation of the predicted profiles. Its formula is defined as follows:
\begin{equation}
{MAE} = \frac{1}{n}\sum_{i=1}^{n} \left| y_i - \hat{y}_i \right|
\end{equation}
Root Mean Squared Error (RMSE) is the arithmetic square root of the mean squared error. Since errors are squared during computation, RMSE assigns higher penalty weights to large prediction deviations (e.g., inaccurate predictions of local temperature extrema). It is suitable for evaluating the robustness of the model in predicting local fluctuations of the profile. Its formula is defined as follows: 
\begin{equation}
{RMSE} = \sqrt{\frac{1}{n}\sum_{i=1}^{n}(y_i - \hat{y}_i)^2}
\end{equation}

Coefficient of Determination ($R^2$) metric quantifies the proportion of variance in the real data explained by the model’s predictions. Its value typically ranges from 0 to 1.For presentation purposes, the $R^2$ values reported in the subsequent tables are multiplied by 100. An $R^2$ value closer to 1 (a value closer to 100 in the tables) indicates a stronger ability of the model to fit the real physical profiles and higher prediction reliability. Its formula is defined as follows:
\begin{equation}
R^2 = 1 - \frac{\sum_{i=1}^{n}(y_i - \hat{y}_i)^2}{\sum_{i=1}^{n}(y_i - \bar{y})^2}
\end{equation}

\subsection{Implementation Details} 
All models in this study were implemented, trained, and validated using the PyTorch ~\cite{paszke2019pytorch} deep learning framework. To accelerate large-scale tensor operations and multi-spectral feature extraction, all model training was conducted on a computing node equipped with a single NVIDIA GeForce RTX 4090 graphics processing unit (GPU).
For network weight optimization, the Adam~\cite{kingma2014adam} (Adaptive Moment Estimation) optimizer was adopted for its adaptive momentum estimation. The initial learning rate (LR) was set to 0.001. To further improve convergence accuracy in the late training stage, a dynamic learning rate scheduling strategy based on validation loss was employed. If the validation mean squared error (MSE) did not decrease significantly for 30 consecutive epochs, the learning rate was automatically decayed to 50\% of its current value. The total number of training epochs was set to 400, and the model weights with the minimum validation loss were dynamically saved after each epoch for final testing.
Considering the risk of gradient explosion caused by high-dimensional plasma input features and the Transformer self-attention mechanism during backpropagation, global gradient clipping was strictly implemented before the optimizer’s gradient update step. This strategy effectively stabilized the training trajectory and avoided unphysical extreme spikes in the predicted temperature profiles. For the data loading configuration, the batch size of the training set was set to 16. The training data were randomly shuffled at the beginning of each epoch, and the last incomplete batch was dropped to ensure stable statistical foundations for the batch normalization layers when computing mean and variance.

\subsection{Comparison with Other Models} 

To verify the effectiveness and superiority of the proposed model, we conduct comparative experiments with a series of mainstream baseline models on the identical held-out test set, including plain MLP, ConvMLP, CNN-Fusion, ResNet variants and Swin Transformer series. The evaluation metrics consist of Mean Absolute Error (MAE, $\downarrow$), Root Mean Square Error (RMSE, $\downarrow$), and coefficient of determination ($R^2$, $\uparrow$). All quantitative results are summarized in Table~\ref{tab:comparison}.

As shown in Table~\ref{tab:comparison}, the proposed model achieves state-of-the-art performance and surpasses all baseline methods consistently on every metric. Compared with plain MLP, our model reduces MAE by 40.1\% (from 15.97 to 9.57) and RMSE by 30.6\% (from 25.88 to 17.96), while lifting the $R^2$ from 93.62\% to 96.92\%. More significant gains are observed against ConvMLP: MAE drops by 45.8\% (from 17.65 to 9.57) and RMSE declines by 36.4\% (from 28.22 to 17.96), accompanied by an increase of $R^2$ from 92.41\% to 96.92\%. When compared with popular visual backbones including CNN-Fusion, ResNet and Swin models, our method still maintains a dominant performance lead. The best-performing baseline Swin-S only attains MAE = 13.55, RMSE = 23.53 and $R^2$ = 94.73\%, while our model cuts the two error metrics by 29.4\% and 23.7\% respectively, and raises the $R^2$ value by 2.19 percentage points.

The specially designed multimodal Transformer fusion architecture, lightweight feature attention module, and physics-guided joint loss function of our model can efficiently fuse zero-dimensional macroscopic parameters and two-dimensional spectral diagnostic data. It accurately captures global radial dependencies and core-edge coupling patterns of temperature profiles, greatly boosting overall prediction precision and morphological fitting performance, particularly for hard-to-fit regions such as the high-temperature core and boundary confinement zones. 
Overall, cross-comparison results with multiple baselines fully validate the remarkable effectiveness and superiority of our proposed model in the joint prediction of plasma electron and ion temperature profiles.

\begin{table}
\centering
\caption{Comparison with Other Models}
\label {tab:comparison}
\begin{tabular}{l|ccc}
\toprule
\textbf{Model} & \textbf{MAE}$\downarrow$ & \textbf{RMSE}$\downarrow$ & \textbf{R}$^2$$\uparrow$ \\ 
\midrule
ConvMLP~\cite{li2023convmlp}    & 17.65 & 28.22 & 92.41 \\
Plain MLP~\cite{tolstikhin2021mlp}   & 15.97 & 25.88 & 93.62 \\
CNN-Fusion~\cite{simonyan2014very}  & 14.90 & 25.57 & 93,77 \\
ResNet-50~\cite{he2016identity}   & 13.15 & 24.16 & 94.44 \\
ResNet-101~\cite{he2016identity}  & 15.31 & 26.43 & 93.35 \\
Swin-T~\cite{liu2021swin}      & 16.30 & 26.75 & 93.19 \\
Swin-S~\cite{liu2021swin}      & 13.55 & 23.53 & 94.73 \\
Ours        & \textbf{9.57} & \textbf{17.96} & \textbf{96.92} \\
\bottomrule
\end{tabular}
\end{table}


\subsection{Ablation Study} 

To further verify the effectiveness of the proposed Transformer-based multimodal prediction architecture (Transformer Predictor) and the Physics-Guided Loss function, a series of ablation experiments are conducted under identical datasets and experimental settings. From the perspectives of model architecture and multimodal fusion, as well as the physical constraint mechanism, this section decouples each core component and independently evaluates its contribution to the final prediction performance. 

In the plasma temperature profile prediction task, the representation capability of features and the modeling ability of spatial dependencies directly determine the upper bound of prediction performance. To validate the necessity of each architectural component, several model variants are designed for comparison: 

\textit{- w/o Spec Data (Only Macroscopic Parameters)}: Based on the proposed Transformer architecture, the 2D spectral encoder (Spec 2D Encoder) is removed, and only 0D macroscopic parameters are used for prediction. This setting verifies the performance gain brought by introducing internal radiation spectral data. 

\textit{- w/o Macro Attention (Without Feature Attention)}: Multimodal inputs are retained, while the Macro Feature Attention module is eliminated. Macroscopic parameters and spectral features are directly concatenated to evaluate the role of the feature-level dynamic weighting mechanism. 

\textit{- w/o Transformer Backbone (Backbone Replacement)}: The multimodal feature fusion mechanism is preserved, whereas the spatial Transformer backbone is replaced with a conventional Convolutional MLP (ConvMLP). This validates the superiority of self-attention in modeling global spatial dependencies of profiles. 

\textit{- Full Model}: The complete model proposed in this paper.

\begin{table}
\centering
\caption{Model Ablation Experiment Results}
\label{tab:model_ablation}
\resizebox{\linewidth}{!}{ 
\begin{tabular}{l|cccccc}
\toprule 
\textbf{Model} & \textbf{Spec} & \textbf{Macro Attn} & \textbf{Former} & \textbf{MAE}$\downarrow$ & \textbf{RMSE}$\downarrow$ & \textbf{R}$^2$$\uparrow$ \\
\midrule 
Baseline (Plain MLP) & $\times$ & $\times$ & $\times$ & 15.97 & 25.88 & 93.62 \\
\textit{w/o} Spec Data & $\times$ & $\checkmark$ & $\checkmark$ & 10.70 & 20.41 & 96.03 \\
\textit{w/o} Macro Attention & $\checkmark$ & $\times$ & $\checkmark$ & 10.18 & 18.47 & 96.75 \\
\textit{w/o} Transformer & $\checkmark$ & $\checkmark$ & $\times$ & 12.32 & 20.95 & 95.82 \\
Full Model & $\checkmark$ & $\checkmark$ & $\checkmark$ & \textbf{9.57} & \textbf{17.96} & \textbf{96.92} \\
\bottomrule 
\end{tabular}}
\end{table}

As shown in Table~\ref{tab:model_ablation}, the full model achieves the optimal performance across all evaluation metrics. The detailed analyses are as follows:
\textit{1). Necessity of multimodal fusion}: Comparing the variant w/o Spec Data with the full model, all metrics are substantially improved after introducing 2D spectral data. This indicates that although 0D macroscopic engineering parameters can characterize the global state of plasma, they cannot describe the internal evolutionary details. In contrast, 2D spectral data effectively compensates for such information deficiency and greatly enhances the fitting capability of the model.
\textit{2). Effectiveness of feature weighting}: The performance degradation of the w/o Macro Attention variant verifies that not all macroscopic parameters contribute equally to temperature profile prediction. Through the attention mechanism, the Macro Feature Attention module enables the model to adaptively focus on the key engineering parameters that dominate the current discharge state, thereby reducing noise interference caused by redundant features.
\textit{3). Superiority of spatial dependency modeling}: The adoption of the Transformer backbone yields performance gains compared with w/o Transformer Backbone (CNN/MLP structure based on local receptive fields). Strong non-local transport coupling effects exist along the radial direction of plasma temperature profiles. The self-attention mechanism can effectively capture such long-range spatial dependencies, making the predicted temperature profiles more consistent with the real physical distribution.

Pure data-driven deep learning models tend to produce prediction results that violate fundamental physical laws when forecasting plasma profiles. This paper proposes a physics-guided loss function that incorporates core-region weighting, a gradient-smoothness penalty, and a monotonicity penalty. To verify the contribution of each physical prior, several ablation variants are designed as follows:

\textit{- Standard MSE}: All physical penalty terms are completely removed, and the loss degenerates to a pure data-driven standard MSE loss.

\textit{- w/o Core Weighting}: The additional weight constraint for the core region is eliminated, and equal weights are assigned to the entire profile.

\textit{- w/o Gradient Penalty}: The weight of the first-order derivative penalty term is set to 0, without constraining the consistency of adjacent spatial points.

\textit{- w/o Monotonicity Penalty}: The weight of the monotonicity penalty term is set to 0, and the temperature profile is no longer constrained to decrease monotonically from the core to the edge region. 
The results in Table~\ref{tab:loss_ablation} consistently demonstrate that the introduction of physical priors is crucial for improving the physical consistency of the model.

\begin{table} 
  \centering 
  \caption{Physics-Guided Loss Ablation Experiment Results} 
  \label{tab:loss_ablation} 
  \resizebox{\linewidth}{!}{
  \begin{tabular}{l|cccccc} 
    \toprule
    \textbf{Model Variant} & \textbf{Core Wt.} & \textbf{Grad Pen.} & \textbf{Mono Pen.} & \textbf{MAE}$\downarrow$ & \textbf{RMSE}$\downarrow$ & \textbf{R}$^2$$\uparrow$ \\ 
    \midrule
    Standard MSE & $\times$ & $\times$ & $\times$ & 11.04 & 21.58 & 95.56 \\ 
    + Core Weighting & $\checkmark$ & $\times$ & $\times$ & 10.47 & 19.50 & 96.37 \\ 
    + Gradient Penalty & $\checkmark$ & $\checkmark$ & $\times$ & 10.84 & 20.02 & 96.18 \\ 
    Full Physics Loss & $\checkmark$ & $\checkmark$ & $\checkmark$ & \textbf{9.57} & \textbf{17.96} & \textbf{96.92} \\ 
    \bottomrule
  \end{tabular}} 
\end{table}

\begin{figure}
  \centering
  \begin{minipage}{0.48\textwidth}
    \centering
    \includegraphics[width=\linewidth]{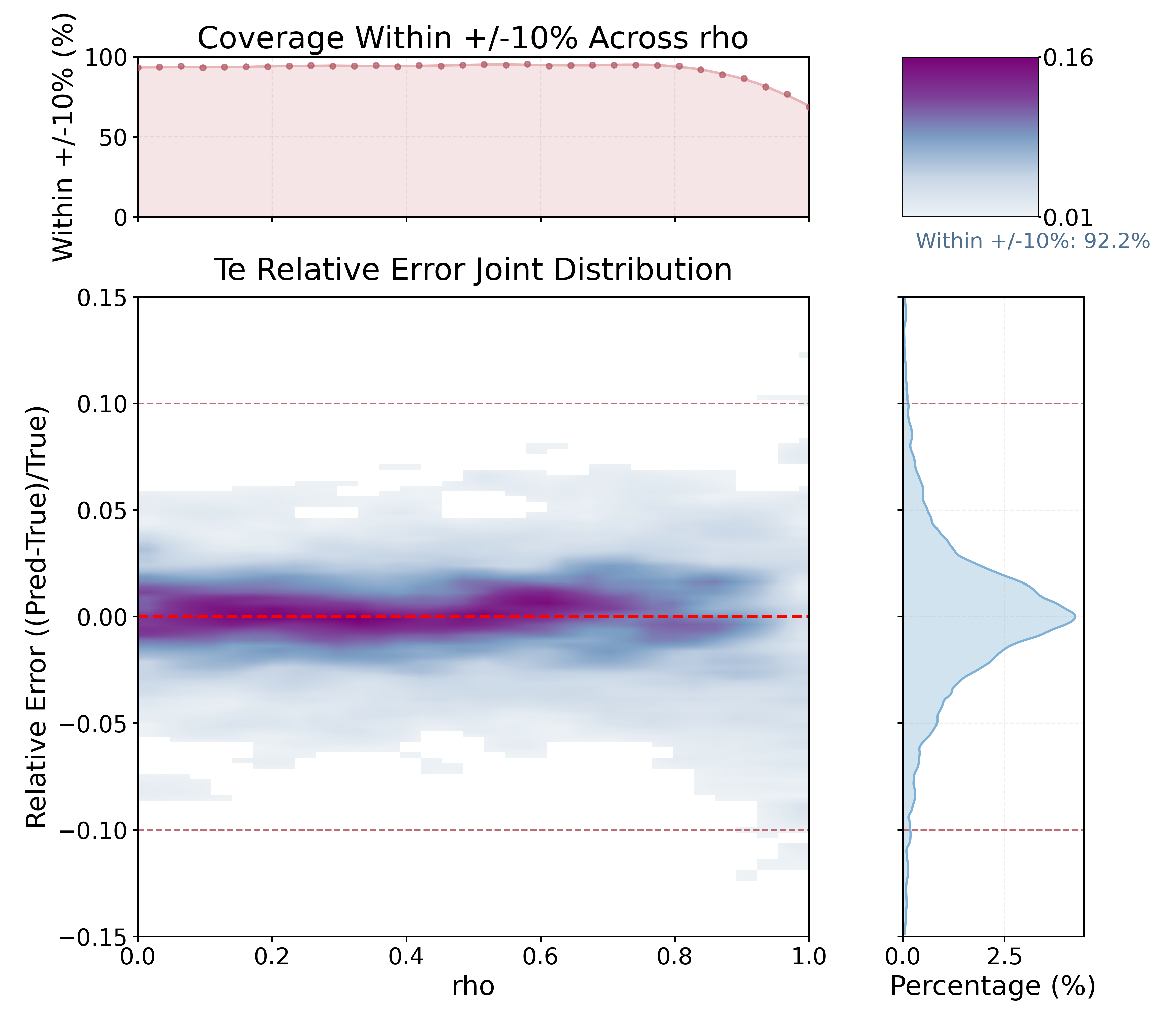}
    \caption{Relative error distribution of predicted electron temperature $T_e$.}
    \label{fig:te_error}
  \end{minipage}
  \hfill
  \begin{minipage}{0.48\textwidth}
    \centering
    \includegraphics[width=\linewidth]{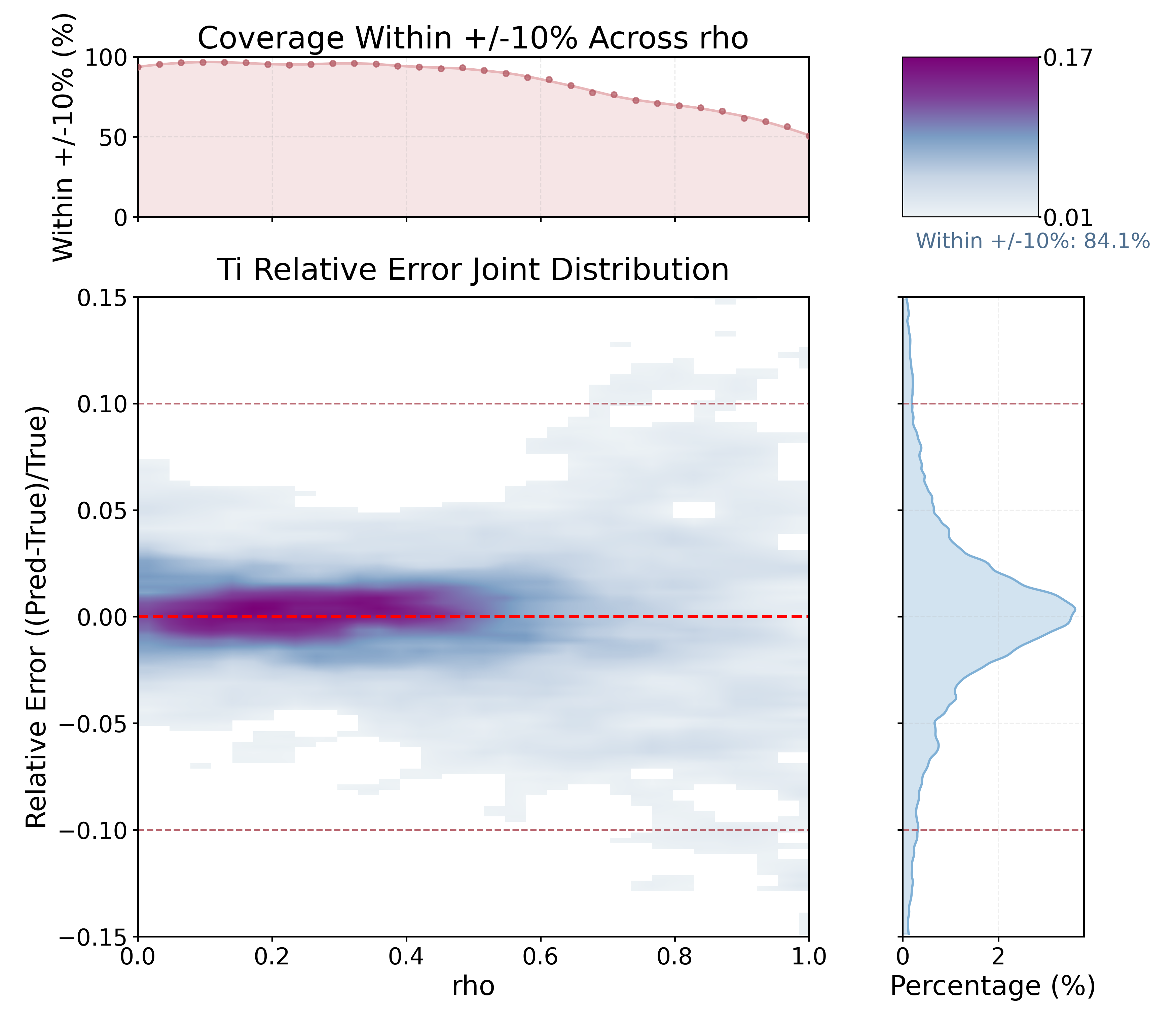}
    \caption {Relative error distribution of predicted ion temperature $T_i$.}
    \label{fig:ti_error}
  \end{minipage}
\end{figure}

\begin{figure*}
\centering
\includegraphics[width=\textwidth]{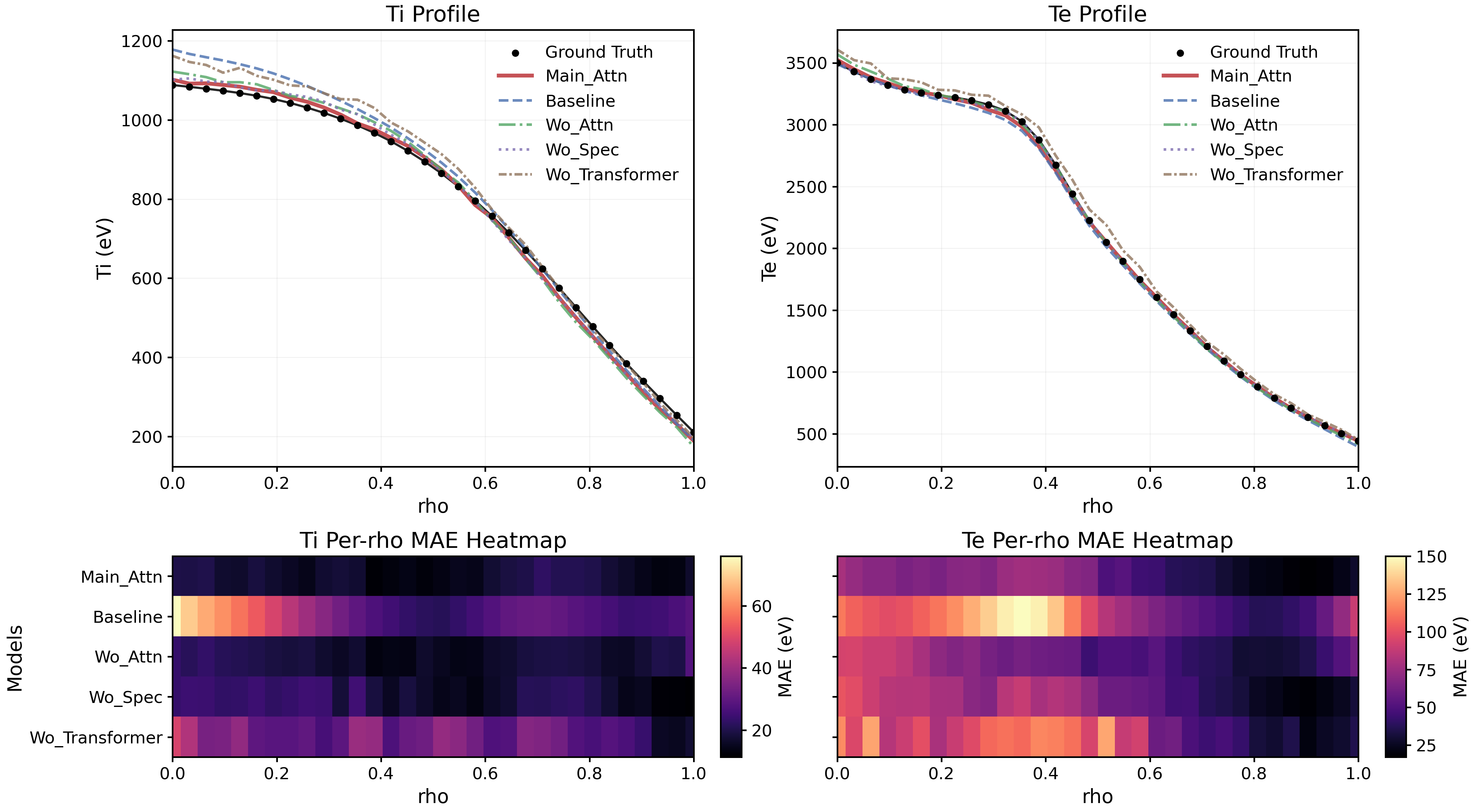}
\caption{Single-shot prediction comparison of different model variants. The top row shows predicted $T_i$ and $T_e$ profiles versus ground truth, and the bottom row presents the corresponding per-rho MAE heatmaps to visualize spatial error distribution.}
\label{fig:single_shot}
\end{figure*}

\begin{table}
  \centering
  \caption{Single Shot Comparison}
  \label{tab:Shot Comparison}
  \resizebox{\linewidth}{!}{
    \begin{tabular}{lccc}
      \toprule
      Model & MAE$\downarrow$ & RMSE$\downarrow$ & R$^2$$\uparrow$ \\
      \midrule
      Baseline  & 12.23 & 18.42 & 89.43 \\
      w/o Spec Data & 7.50 & 14.19 & 93.97 \\
      w/o Macro Attention & 7.49 & 12.14 & 95.41 \\
      w/o Transformer Backbone & 10.55 & 14.74 & 93.23 \\
      Full Model & \textbf{6.55} & \textbf{11.39} & \textbf{95.56} \\
      \bottomrule
    \end{tabular}
  }
\end{table}

\begin{figure*}[!htp]
    \centering
    \includegraphics[width=0.7\linewidth]{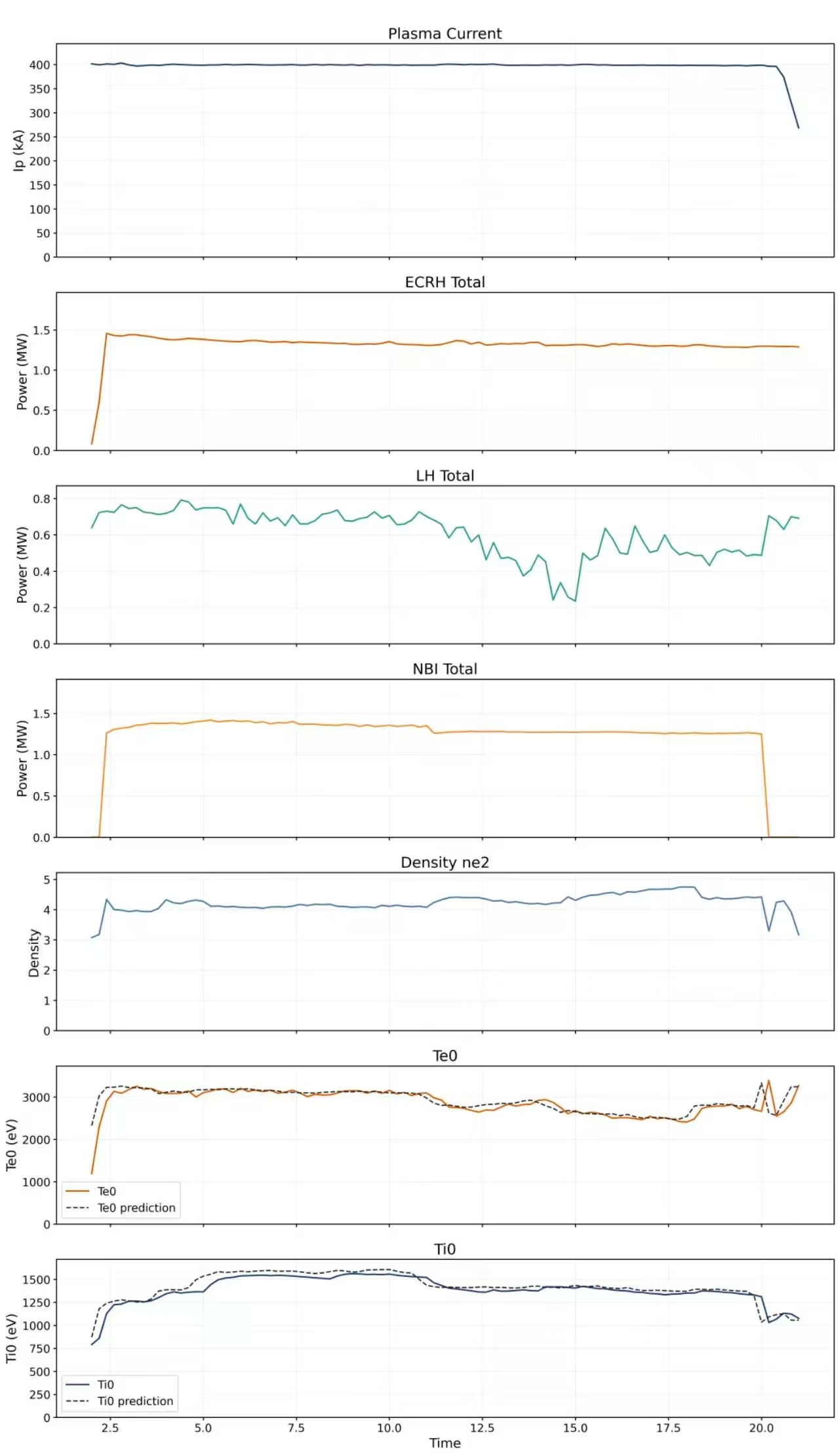}
    \caption{
    Temporal evolution of representative model inputs and central-temperature predictions for a single EAST discharge.
    From top to bottom, the panels show the plasma current $I_p$, total ECRH power, total LH power, total NBI power, electron density $n_{e2}$, central electron temperature $T_{e0}$, and central ion temperature $T_{i0}$.
    In the two bottom panels, the solid curves represent the measured temperatures and the dashed curves denote the model predictions.
    }
    \label{fig:temporal_evolution}
\end{figure*}

\begin{figure*}[htbp]
    \centering
    \begin{minipage}{0.67\textwidth}
        \centering
        \includegraphics[width=\linewidth]{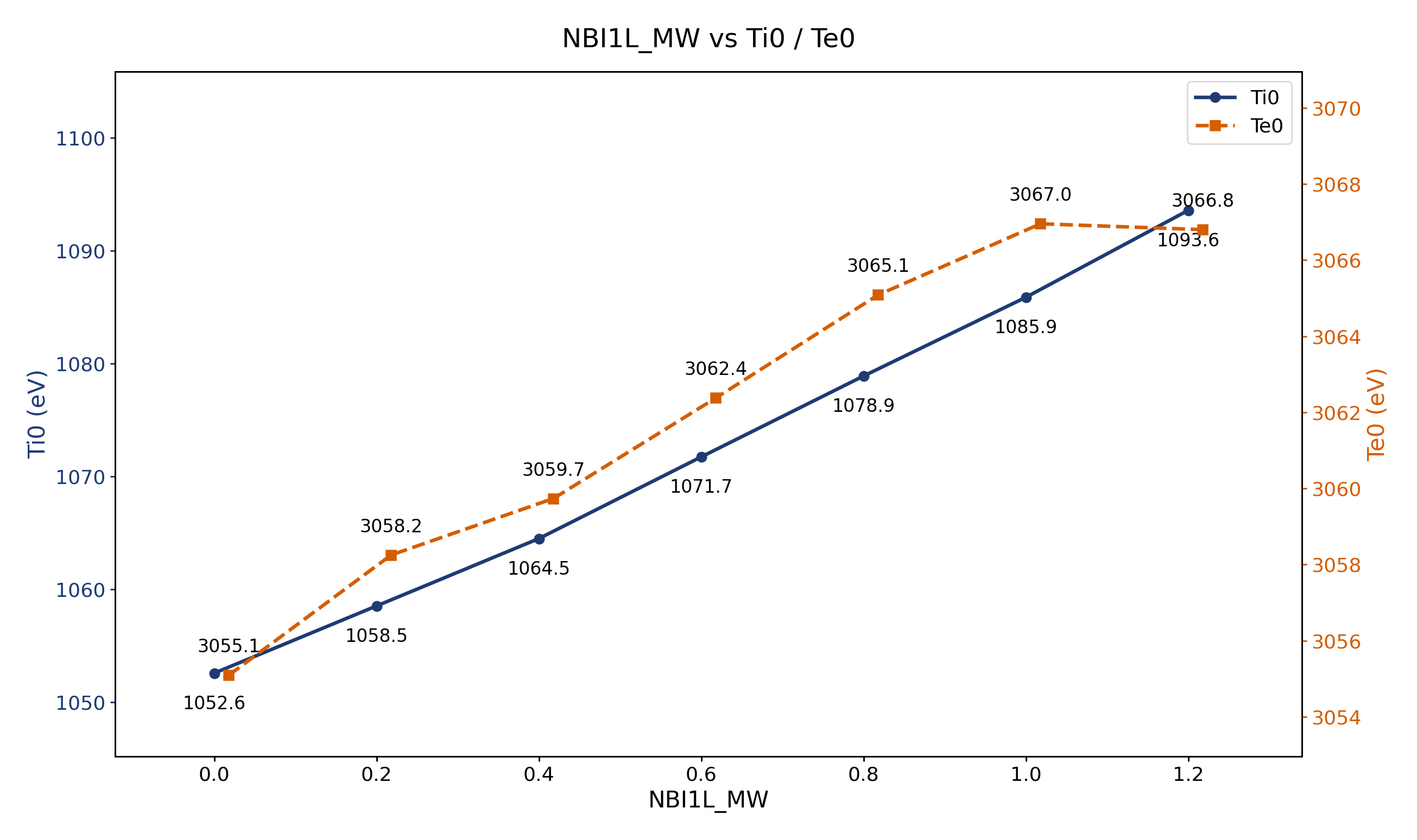} 
        \subcaption{NBI heating power sweep}
        \label{subfig:plot1}
    \end{minipage}
    \hfill
    \begin{minipage}{0.67\textwidth}
        \centering
        \includegraphics[width=\linewidth]{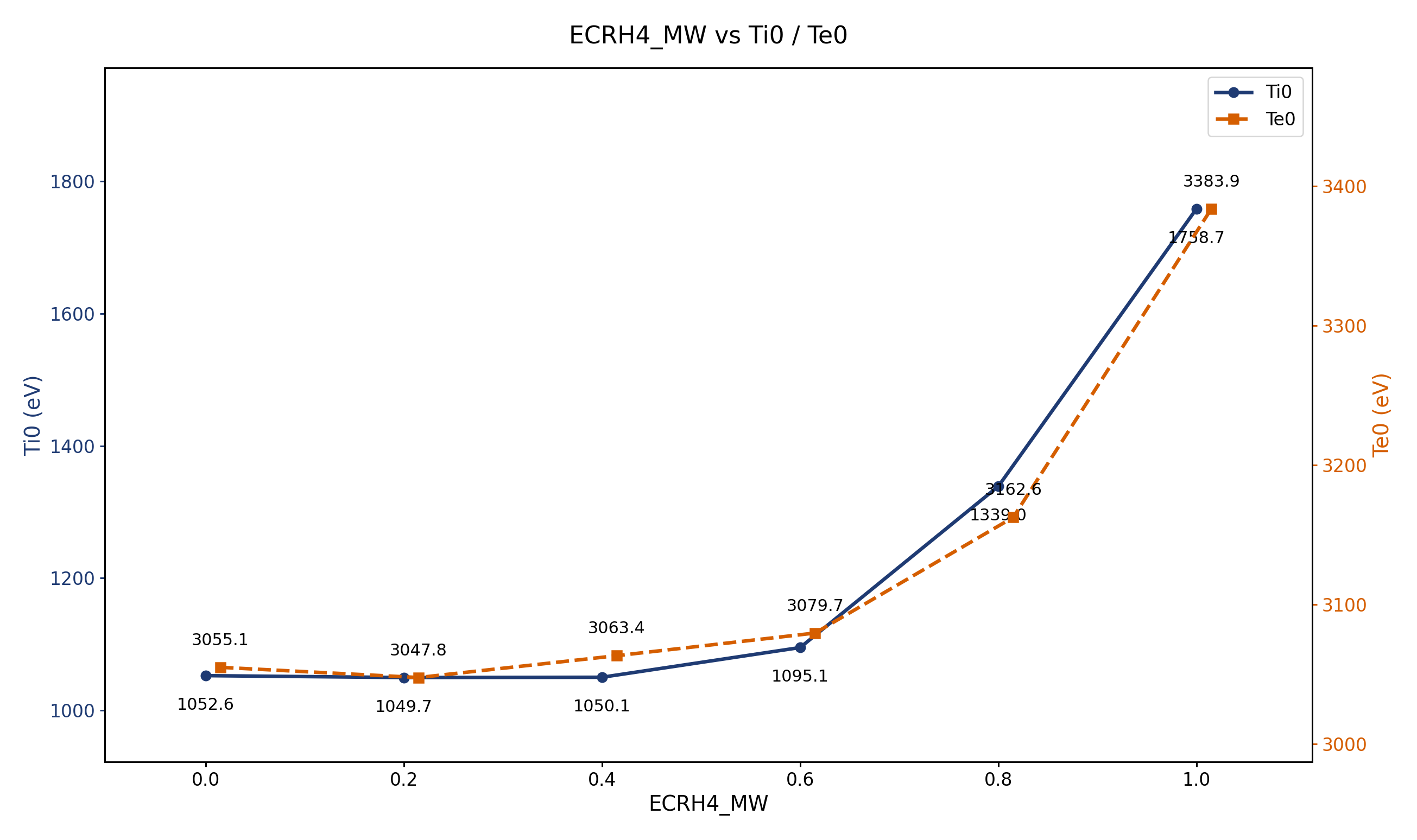}
        \subcaption{ECRH heating power sweep}
        \label{subfig:plot2}
    \end{minipage}
    \vspace{0.6em} 
    \begin{minipage}{0.67\textwidth}
        \centering
        \includegraphics[width=\linewidth]{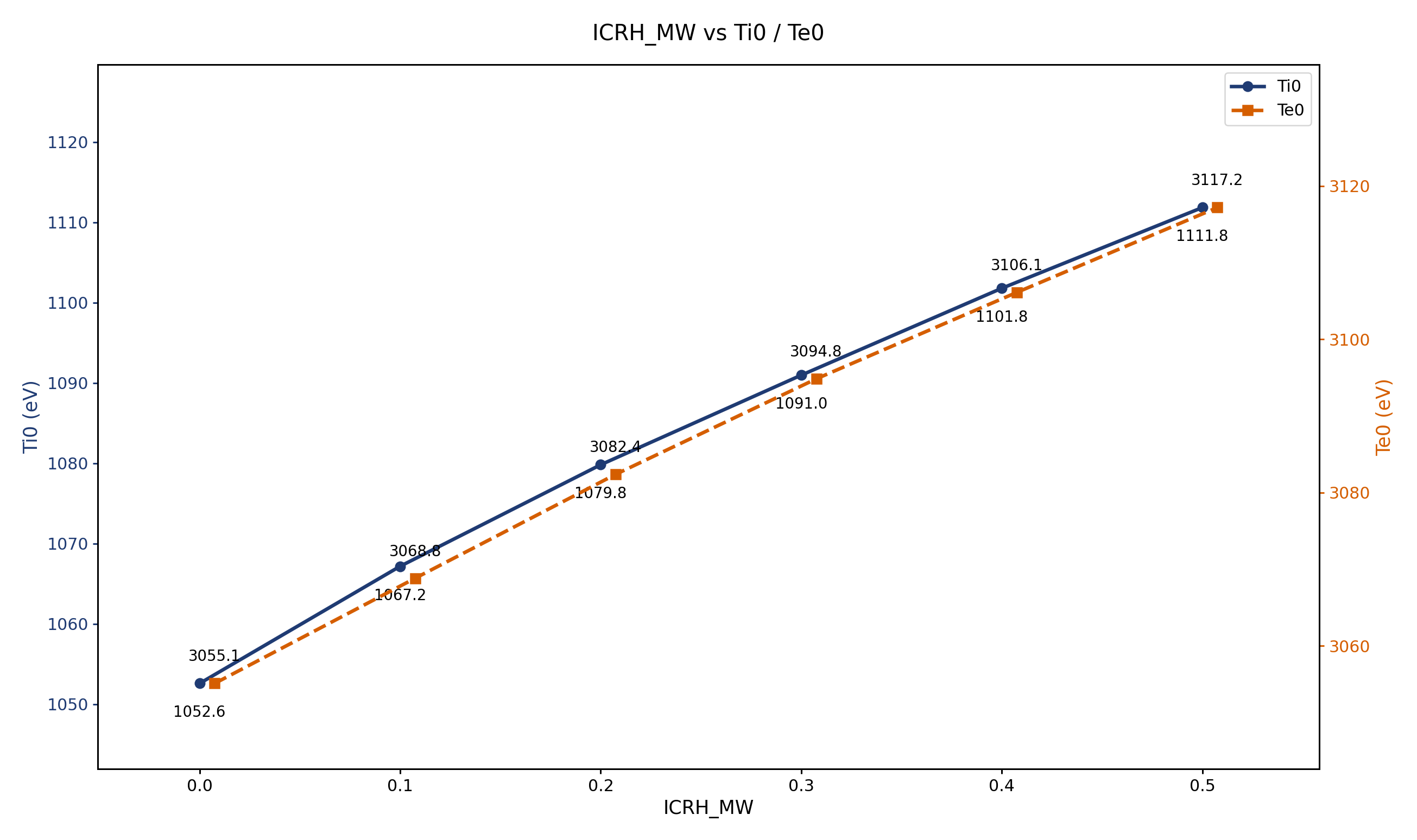}
        \subcaption{ICRH heating power sweep}
        \label{subfig:plot3}
    \end{minipage}
    \caption{Core ion temperature $T_{i0}$ and core electron temperature $T_{e0}$ variations under continuous power scanning of three heating systems: (a) NBI heating power sweep; (b) ECRH heating power sweep; (c) ICRH heating power sweep.}
    \label{fig:all_profile}
    \label{fig:sensitivity}
\end{figure*}

\subsection{Visualization} 
To further analyze the prediction performance beyond the aggregated metrics (MAE, RMSE, $R^2$), we visualize the relative error distributions of the predicted electron temperature ($T_e$) and ion temperature ($T_i$) profiles on the test set. This visualization allows us to examine not only the overall prediction accuracy, but also the spatial stability and potential systematic bias of the model.

\noindent \textbf{1). Visualization of prediction error distribution for temperature profiles.} 
As shown in Figure~\ref{fig:te_error} and Figure~\ref{fig:ti_error}, the relative errors of both $T_i$ and $T_e$ are mainly concentrated around the zero-error line, indicating that the model can reconstruct the temperature profiles with good overall accuracy across most radial locations. The $T_i$ error distribution is more compact, with its high-density region remaining close to zero over the entire $\rho$ range, suggesting better prediction stability for $T_i$. In comparison, the $T_e$ error distribution is broader, and its spread increases at larger $\rho$, implying that electron temperature prediction becomes more challenging in the edge region. Overall, no strong systematic bias is observed, although local fluctuations remain in the outer radial region.

\noindent \textbf{2). Visualization of model prediction performance under a single discharge.} 
To further demonstrate the advantages of the proposed full model and its components in a concrete scenario, we present a detailed comparison of prediction results for a single representative discharge (single shot). The quantitative results for this single shot are summarized in Table~\ref{tab:Shot Comparison}. The proposed Full Model achieves the lowest MAE (6.55) and RMSE (11.39), and the highest $R^2$ (95.56\%), outperforming all other model variants, including the baseline and the ablated versions. This confirms that the performance gains observed in the overall dataset also hold true at the individual shot level.

Figure~\ref{fig:single_shot} compares the predicted temperature profiles from different models with the ground truth for both $T_i$ and $T_e$, along with per-rho MAE heatmaps to highlight spatial error distribution. In contrast, the proposed Full Model produces the profiles that most closely match the ground truth for both $T_i$ and $T_e$. It accurately captures the sharp gradients in the edge region, maintains the correct peak temperature in the core, and avoids the unphysical oscillations or excessive smoothing seen in other models. The per-rho MAE heatmaps confirm that the Full Model consistently maintains the lowest error levels across almost the entire radial range, with only minor fluctuations near the very edge. 
This single-shot comparison vividly demonstrates the effectiveness of each component in the proposed model. It further validates that the proposed approach is not only statistically superior but also robust and reliable for practical plasma physics applications.

\noindent \textbf{3). Visualization of Temporal Evolution in a Representative Discharge} 
To further evaluate the temporal consistency of the proposed model, Figure~\ref{fig:temporal_evolution} presents the evolution of several representative macroscopic diagnostic inputs and the corresponding central electron and ion temperatures during a single discharge. The selected input signals include the plasma current $I_p$, total electron cyclotron resonance heating power (ECRH), total lower-hybrid heating power (LH), total neutral beam injection power (NBI), and electron density $n_{e2}$. These quantities characterize the plasma equilibrium, external heating conditions, and density evolution throughout the discharge.

During most of the flat-top phase, the plasma current remains approximately constant, while the auxiliary heating powers and electron density exhibit different levels of temporal variation. In particular, the LH power shows relatively pronounced fluctuations, whereas the ECRH and NBI powers remain comparatively stable over a large portion of the discharge. Near the end of the discharge, the reduction or termination of the heating inputs is accompanied by noticeable variations in the plasma current, density, and temperature signals.

The bottom two panels compare the predicted and measured central electron temperature $T_{e0}$ and central ion temperature $T_{i0}$. The predicted curves closely follow the measured temporal evolution over most of the discharge, including the initial temperature rise, the quasi-steady flat-top phase, the gradual variations associated with changes in the heating inputs, and the rapid fluctuations near the discharge termination. These results demonstrate that the proposed model is capable of capturing not only the radial structure of the temperature profiles, but also their temporal response to evolving plasma conditions.

Motivated by the observed temperature responses under time-varying heating conditions, we further perform controlled power-scan experiments to quantitatively examine the sensitivity of the predicted central temperatures to different auxiliary heating sources.

\noindent \textbf{4). Visualization of input power sensitivity analysis.} 
Figure~\ref{fig:sensitivity} presents the results of the input power sensitivity analysis for a single discharge (shot), aiming to validate the model's response sensitivity and physical consistency with different heating methods. We scanned the power inputs of Neutral Beam Injection (NBI), Electron Cyclotron Resonance Heating (ECRH), and Ion Cyclotron Resonance Heating (ICRH), and monitored the trends in the core ion temperature ($T_{i0}$) and core electron temperature ($T_{e0}$).

The results indicate that the model can accurately capture the specific physical effects of different heating sources:
\begin{itemize}
    \item \textbf{NBI power scan}: As the NBI power increases, $T_{i0}$ and $T_{e0}$ show a significant linear growth.
    \item \textbf{ECRH power scan}: $T_{e0}$ exhibits a strong positive correlation with increasing power, reflecting the direct heating effect of electron cyclotron waves on electrons; meanwhile, $T_{i0}$ also increases steadily with the rise in electron temperature through the energy equipartition mechanism.
    \item \textbf{ICRH power scan}: Both $T_{i0}$ and $T_{e0}$ increase significantly with the increase in radio frequency power, reflecting the dual heating efficiency of ICRH on ions and electrons under specific heating schemes.
\end{itemize}

Overall, these trends are highly consistent with the basic laws of tokamak plasma physics, further proving that the proposed model is not only accurate in numerical prediction but also has good physical interpretability, enabling a correct understanding of the influence of different external driving sources on the plasma profiles.


\section{Conclusion} \label{sec::Conclusion}
In this work, a multimodal fusion attention neural network is proposed for the joint reconstruction of ion temperature and electron temperature radial profiles on the EAST tokamak. Aiming at the inherent limitations of traditional diagnostic measurements and conventional physics-based transport simulations, the developed model effectively addresses the issues of restricted measurement conditions, signal degradation during long-pulse discharges, high computational cost, and error accumulation in conventional methods. Different from existing data-driven approaches that only rely on spectral data and independently predict single-temperature profiles, the proposed framework integrates global macroscopic diagnostic parameters and high-dimensional spectral information to achieve simultaneous $T_i, T_e$ prediction. By introducing an adaptive attention module and convolutional feature extraction, the model adaptively weights valid macroscopic features, suppresses noise interference, and captures fine-grained spectral characteristics, while multi-head self-attention mechanisms further enhance cross-modal feature interaction and nonlinear correlation mining. Benefiting from the complementary properties of multimodal inputs, the established model realizes robust profile reconstruction even under scenarios with missing spectral measurements, significantly improving the generalization and practical adaptability for real tokamak operating conditions. Qualitative and quantitative results demonstrate that the proposed method can provide accurate and reliable spatiotemporal distributions of plasma temperature profiles, which is critical for the analysis of plasma confinement performance, turbulent transport behavior, and magnetohydrodynamic stability.

In future work, the model can be further extended to incorporate more diagnostic signals and pre-trained spectral foundation models to achieve higher-performance, long-timescale plasma profile prediction, providing effective technical support for the steady-state operation and intelligent control of EAST long-pulse plasmas.

\small{ 
\bibliographystyle{IEEEtran}
\bibliography{reference}
}

\end{document}